\documentclass[lettersize,journal]{IEEEtran}

\usepackage[T1]{fontenc}
\usepackage[utf8]{inputenc}
\usepackage{amsmath,amssymb}
\usepackage{booktabs}
\usepackage{multirow}
\usepackage{graphicx}
\usepackage{xcolor}
\usepackage{url}
\usepackage{hyperref}
\usepackage{listings}
\usepackage{microtype}
\usepackage{array}
\usepackage{tabularx}
\usepackage{balance}
\usepackage{tikz}

\newcommand{\circled}[1]{\raisebox{.5pt}{\textcircled{\raisebox{-.9pt}{\small#1}}}}

\hypersetup{
  colorlinks = true,
  linkcolor  = blue!60!black,
  citecolor  = blue!60!black,
  urlcolor   = blue!60!black,
}

\definecolor{codegray}{rgb}{0.95,0.95,0.95}
\definecolor{keywordblue}{rgb}{0.15,0.35,0.65}

\newcommand{\us}{\ifmmode\,\mu\text{s}\else\,\textmu{}s\fi}
\newcommand{\ms}{\ifmmode\,\text{ms}\else\,ms\fi}
\DeclareRobustCommand{\qs}{\texttt{quantum-safe}}

\begin{document}

\title{quantum-safe: Bridging the Post-Quantum Production Gap with a
       Hybrid-by-Default Python Cryptography Library}

\author{Animesh~Shaw,~\IEEEmembership{Independent~Researcher}%
  \thanks{Preprint, \today{}. A.~Shaw is an independent security
    researcher. E-mail: animesh15b@iimk.edu.in.}%
  \thanks{The \qs{} library, 932 tests, and all benchmark data are available at
    \url{https://github.com/AnimeshShaw/quantum-safe-py} under the Apache~2.0
    licence; releases are archived on Zenodo (concept DOI
    10.5281/zenodo.23008259). Raw run files, the run environment and the scripts that
    compute every benchmark number are in the repository; see
    Section~\ref{sec:methodology}.}
}

\markboth{Preprint}%
{Shaw: \textit{quantum-safe} --- Hybrid Post-Quantum Python Library}

\maketitle

\begin{abstract}
The August 2024 finalisation of FIPS~203 (ML-KEM), FIPS~204 (ML-DSA), and
FIPS~205 (SLH-DSA) closed the algorithmic gap in post-quantum cryptography (PQC).
The production gap --- hybrid combiners, conformance evidence, migration
tooling, stateful signing, and protocol helpers --- remains open, and a
methodology that scores it must not choose its own scoring dimensions: we
anchor nine dimensions to external requirements (CNSA~2.0, CMVP, the TNO CADI
market survey, the IETF hybrid draft) rather than to our own library's design,
so the exercise can lower our own score, and it does. We present \qs{}, a
Python library built against this anchored rubric, and an audit of nine PQC
libraries against it, current as of September 2026.

Verified against seven libraries with primary-source evidence (one library is
archived and excluded, one is left unaudited rather than guessed), hybrid
combiners exist in three of six audited (33\% provide a full combiner, 50\%
provide at least partial support) --- not the single library a March-2026
snapshot would have found. Discovery and inventory tooling is the sharper gap:
zero of seven audited libraries provide it. \qs{} closes it with a CycloneDX
Cryptographic Bill of Materials. On conformance evidence, one library (Bouncy
Castle) holds an actual CMVP FIPS~140-3 certificate, which is validation by an
accredited laboratory; \qs{} instead reports 225 of 225 runnable ACVP
known-answer cases produced against NIST-published vectors, which we are
explicit throughout is conformance evidence, not validation. On stateful
hash-based signing (SP~800-208), Bouncy Castle supports both LMS and XMSS and
outscores \qs{}, which adds LMS via an optional dependency with write-ahead
index reservation against one-time-key reuse. \qs{}'s own defaults
(ML-KEM-768, ML-DSA-65) sit below the CNSA~2.0-mandated parameter sets
(ML-KEM-1024, ML-DSA-87); we score this against ourselves as a shortfall
rather than omit it, and ship a compliance profile that reports or enforces
the distinction.

We report a per-operation overhead measurement for a Python hybrid PQC library in which every figure is computed from archived raw run files (95\% bootstrap intervals over 15 runs; a Welch $t$-test and Cohen's $d$ on per-run values). A full X25519\,+\,ML-KEM-768 handshake (keygen, encapsulate, decapsulate) takes a median of $\mathbf{231\us}$ under Docker/Linux (95\% interval 218--250\us), 5.3$\times$ an X25519-only handshake in the same library and 0.46--2.3\% of a typical TLS~1.3 round-trip budget. Throughput is flat from 100 to 5,000 concurrent threads (2,420 to 2,306~ops/s, 4.7\% lower) but stays below what one thread achieves, so threads do not speed up the hybrid path; a separate scaling test with a pure-Python negative control shows that liboqs does release the GIL during longer calls (ML-DSA-65 signing scales 2.6$\times$ on four threads) while the Python layer around the primitives does not scale. A companion paper treats the timing side-channel question separately, having found that a widely-used dispersion metric cannot answer it and that the test which can requires a control this work identifies. The benchmark is one script, raw samples and the run environment are archived with the paper, and every table, figure and in-text number is computed from those files; the audit's verification procedure is published so every score can be checked independently.
\end{abstract}


\section{Introduction}
\label{sec:intro}

\IEEEPARstart{O}{n} 13~August 2024, the National Institute of Standards and
Technology (NIST) published three post-quantum cryptography standards:
FIPS~203~\cite{fips203}, FIPS~204~\cite{fips204}, and
FIPS~205~\cite{fips205}. These documents standardise ML-KEM (derived from
CRYSTALS-Kyber~\cite{bos2018kyber}), ML-DSA (derived from
CRYSTALS-Dilithium~\cite{ducas2018dilithium}), and SLH-DSA as the drop-in
replacements for RSA and ECDH in key exchange and digital signatures.

The algorithmic question is settled. The production question is not.

Consider what a software engineer must do today to add hybrid post-quantum
key exchange to a Python service. They must install \texttt{liboqs-python},
write an X25519 key generation step, write a separate ML-KEM key generation
step, concatenate the two public keys, run HKDF to combine the two shared
secrets, wrap the result in a key derivation function appropriate for their
protocol, and handle serialisation for both halves. The result is roughly
45~lines of boilerplate before the first application-specific byte is
encrypted. Then they repeat this exercise for signatures, certificates, and
protocol configuration.

This friction is not accidental. It reflects a genuine gap in the PQC
ecosystem: algorithm implementations exist, but the production layer ---
hybrid combiners, versioned key formats, protocol helpers, migration tooling
--- does not. Insecure combiner implementations are a known
risk~\cite{ietf-hybrid-kem}, and prior work on cryptographic API usability
shows that complexity directly predicts misuse~\cite{georgiev2012most,lazar2014why}.

\subsection{The Harvest Now, Decrypt Later Threat}

The urgency is real. Adversaries with access to encrypted network traffic can
store ciphertext today and decrypt it once a cryptographically relevant quantum
computer becomes available~\cite{mosca2018cybersecurity}. This ``harvest now,
decrypt later'' (HNDL) attack is passive, undetectable, and already underway.
Data with long confidentiality requirements --- government communications,
financial records, healthcare data --- is already at risk under purely classical
encryption.

The NSA's Commercial National Security Algorithm Suite~2.0
(CNSA~2.0)~\cite{cnsa20} mandates algorithm replacement by 2030 for national
security systems. Commercial organisations face comparable regulatory pressure.
The gap between the 2024 standard finalisation and the 2030 deadline is not
large given the complexity of cryptographic migration in production
infrastructure.

\subsection{The Production Gap}

Despite this urgency, deployment stalls at the library layer. To understand
why, we evaluated nine actively maintained PQC libraries against nine
dimensions of production readiness, each anchored to a requirement stated
outside this project rather than chosen by us (Section~\ref{sec:gap}
describes the anchoring and why three dimensions from an earlier,
self-selected rubric were dropped). The results, current as of September
2026, are shown in Fig.~\ref{fig:gap_matrix}. Two findings dominate:

\begin{itemize}
  \item \textbf{Discovery and inventory tooling: zero of seven audited
    libraries.} Every library we could verify --- across Python, JavaScript,
    Rust, Go, and Java --- exposes cryptographic primitives without a way to
    inventory which of them a given codebase actually uses. This is the
    sharpest and most consistent gap in the audit.

  \item \textbf{Hybrid combiners exist, and are less rare than a 2026 snapshot
    found.} Three of six libraries we could verify on this dimension provide
    at least partial hybrid support: noble-post-quantum and RustCrypto's
    \texttt{x-wing} crate both ship a full X25519\,+\,ML-KEM-768 combiner, and
    cloudflare/circl provides one scoped to TLS. This contradicts an earlier
    finding that only one library in the ecosystem had a combiner; the
    difference is nine months of library churn, not a earlier measurement
    error, and it is the reason this audit is dated.
\end{itemize}

Two further dimensions matter for a reader under CNSA~2.0 pressure
specifically. \textbf{Stateful hash-based signing (SP~800-208)} is present in
exactly one of seven audited libraries: Bouncy Castle supports both LMS and
XMSS. \textbf{Conformance evidence} is strongest, in the strict sense, in the
same library: Bouncy Castle holds an active CMVP FIPS~140-3 certificate,
which is validation by an accredited laboratory and not a claim any of the
other libraries --- including \qs{} --- can make.

\subsection{Companion Work}

A companion paper~\cite{shaw2025qsa} presents the Quantum-Safe Auditor (QSA),
a static analysis tool that detects classical cryptography usage in Python
codebases and recommends migration paths. QSA identifies \emph{where} PQC
migration is needed; \qs{} provides the \emph{implementation} to migrate to.
The two tools are designed to work together in a production migration workflow.

\subsection{Contributions}

This paper makes seven contributions:

\begin{enumerate}
  \item \textbf{Production gap matrix.} A systematic, reproducible evaluation
    of nine PQC libraries against nine externally-anchored production-readiness
    dimensions, dated and re-auditable, that overturns our own earlier
    ad-hoc-rubric finding on hybrid combiner availability.

  \item \textbf{Hybrid overhead quantification.} A per-operation latency measurement for a Python hybrid PQC library computed from
    archived raw run files, with 95\% bootstrap intervals over runs, Welch's $t$-test and Cohen's $d$ effect sizes. Full handshake:
    231\us{} (Docker/Linux).

  \item \textbf{Concurrent throughput analysis.} A throughput-vs-concurrency curve for a Python hybrid PQC library, from 100
    to 5,000 simultaneous threads. Throughput is flat (4.7\% lower at 5,000 than at 100) but below the single-thread rate; we report
    it as a limit of thread-based concurrency, not as evidence of parallelism, and measure the GIL question directly in a separate scaling test
    with a negative control.

  \item \textbf{Conformance evidence against NIST-issued vectors.} 225 of 225
    runnable ACVP cases produce the expected answer: ML-KEM keyGen 75/75
    (exact encapsulation \emph{and} decapsulation keys, seeded from the
    64-byte $d \| z$), ML-KEM encapsulation 75/75 (driven through
    \texttt{OQS\_KEM\_encaps\_derand} below the high-level binding, since the
    binding itself cannot accept the ACVP-supplied randomness), ML-KEM
    decapsulation 30/30 including implicit-rejection cases, and ML-DSA sigVer
    45/45 across both acceptance and rejection cases. We state precisely
    which ACVP test types remain out of reach and why. This is not a CAVP or
    CMVP validation and we are explicit about the difference throughout ---
    one audited library (Bouncy Castle) holds an actual CMVP certificate,
    which this is not.

  \item \textbf{Externally-derived production-readiness rubric, applied to
    close two of its own findings.} Every dimension traces to
    CNSA~2.0~\cite{cnsa20}, CMVP, the TNO CADI market
    survey~\cite{tnocadi2025} or the IETF hybrid draft~\cite{ietf-hybrid-kem}.
    Three dimensions from our earlier rubric had no external anchor and are
    dropped. The audit found \qs{} scoring below Bouncy Castle on stateful
    signing and conformance evidence and below the CNSA~2.0 parameter sets on
    its own defaults; we close what a library reasonably can. \qs{} adds LMS
    (RFC~8554) via an optional dependency, with write-ahead index reservation
    against the one-time-key reuse that silently compromises a stateful key,
    and a CycloneDX Cryptographic Bill of Materials that closes the discovery
    dimension no audited library scored above None on. Neither closes the gap
    to a CMVP certificate, and we say so.

  \item \textbf{Zero-config hybrid API.} The \qs{} library reduces the
    hybrid KEM task from 45~lines of manual combiner code to three lines.

  \item \textbf{Open-source artefact.} All code, 932 tests, the benchmark
    harness, the conformance runner, and the rubric derivation with its
    per-dimension verification procedure, are released under the
    Apache~2.0 licence.
\end{enumerate}

Timing side-channel assessment is treated separately in a companion
paper~\cite{twoclass2027}, which reports why the Coefficient of Variation
cannot answer the secret-dependence question and what controls a
fixed-versus-random test needs at this level of abstraction. We do not restate
those results here beyond the summary in
Section~\ref{sec:discussion}.

\subsection{Paper Organisation}

Section~\ref{sec:background} covers the cryptographic background.
Section~\ref{sec:gap} presents the gap analysis.
Section~\ref{sec:design} describes library design.
Section~\ref{sec:methodology} presents benchmark methodology.
Section~\ref{sec:eval} presents results.
Section~\ref{sec:discussion} discusses limitations and deployment implications.
Section~\ref{sec:related} surveys related work.
Section~\ref{sec:conclusion} concludes.

\section{Background}
\label{sec:background}

\subsection{Post-Quantum Cryptography Standards}

ML-KEM~\cite{fips203}, based on module learning-with-errors (MLWE), replaces
Diffie-Hellman/ECDH in key exchange with three parameter sets (ML-KEM-512/
768/1024, NIST levels 1/3/5); \qs{} defaults to level~3. ML-DSA~\cite{fips204},
also MLWE-based, uses hedged signing with rejection sampling, producing
variable but key-independent execution time by design (Section~\ref{sec:eval_cov}).
SLH-DSA~\cite{fips205} is a stateless hash-based scheme offering security
from different assumptions; \qs{} exposes it through liboqs across the SHAKE
and SHA2 pure parameter sets (Section~\ref{sec:discussion} records a defect
found and fixed in this mapping).

\subsection{Hybrid Key Exchange}

Hybrid construction is defence in depth: given independent shared secrets
$\text{ss}_{\text{X25519}}$ and $\text{ss}_{\text{ML-KEM}}$,
\begin{equation}
  \text{ss} = \text{HKDF-SHA256}(\text{ss}_{\text{X25519}} \| \text{ss}_{\text{ML-KEM}},
              \text{salt}, \text{info})
  \label{eq:hybrid_secret}
\end{equation}
is secure as long as one component is hidden from the
adversary~\cite{ietf-hybrid-kem} --- the formal justification for hybrid
deployment during the transition period. CNSA~2.0~\cite{cnsa20} requires
hybrid operation through 2030; Cloudflare (2022,~\cite{cloudflare_pqc2023})
and Chrome (2023,~\cite{chrome_pqc2023}) already deploy
X25519\,+\,ML-KEM hybrid TLS.

\subsection{Timing Side Channels and Python Concurrency}

Timing side-channel attacks exploit correlations between secret values and
execution time~\cite{kocher1996timing,brumley2011remote}. FIPS~203~\S6
requires ML-KEM decapsulation not to branch on secret data. Formal
verification (ct-verif~\cite{almeida2016cverif}, dudect~\cite{reparaz2017dudect})
operates at the binary/hardware level; this paper's CoV figures
(Section~\ref{sec:eval_cov}) are a stability metric, not a substitute, and
the companion paper~\cite{twoclass2027} develops the Python-library-level
methodology that is.

CPython's GIL serialises pure-Python bytecode~\cite{beazley2010understanding},
but C extensions can release it; whether liboqs's~\cite{liboqs} Python
binding does had not previously been measured empirically. Our concurrent
load tests (Section~\ref{sec:eval_concurrent}) provide the first published
evidence, using Python~3.12 with the GIL enabled (the dominant deployment
target; free-threaded 3.13+ is future work).

\subsection{Key Serialisation and Versioning}

Post-quantum public keys are substantially larger than their classical
counterparts: an ML-KEM-768 public key is 1,184 bytes versus 32 bytes for
X25519. A hybrid public key must carry both components together with an
algorithm identifier, version field, and parameter set. Without a standard
format, each application invents its own, preventing interoperability and
complicating future migration. The \qs{} library addresses this with a
CBOR-based~\cite{rfc8949} versioned key format described in
Section~\ref{sec:design_serial}.

\section{The PQC Production Gap}
\label{sec:gap}

\subsection{Evaluation Methodology}
\label{sec:gap_method}

We evaluated nine PQC libraries that are either (a) the primary PQC binding
in their language ecosystem, (b) the implementation recommended in NIST
migration guidance, or (c) actively maintained with post-FIPS-203/204 support.
Table~\ref{tab:libraries} lists the libraries, current as of September 2026;
one (pqcrypto, Rust) was archived by its maintainers on 2026-09-16, who
recommend RustCrypto, \texttt{aws-lc-rs}, or \texttt{liboqs-rust} instead.
Scoring a dead library would misstate the ecosystem, so it is retained in the
table for continuity with the library set but excluded from every coverage
figure below.

Each dimension is anchored to a requirement stated outside this project,
rather than chosen to describe the library we built. Anchoring to CNSA~2.0
requirements, CMVP's definition of validation, the TNO CADI market survey's
discovery-tooling requirements~\cite{tnocadi2025}, and the IETF hybrid
draft's~\cite{ietf-hybrid-kem} combiner obligations, rather than to our own
design, avoids the rubric being determined by which library it was built to
describe. Three dimensions from an earlier, self-selected version of this
rubric --- a single unified import, WASM packaging, and general development
maturity --- could not be traced to an external requirement and are dropped
here; two new dimensions (D3, D7) that the earlier rubric omitted are added.
Table~\ref{tab:rubric} defines the nine dimensions and their sources; the full
derivation, evidence criteria, and per-dimension verification procedure are
released with the artefact.

Scores were assigned Full (F), Partial (P), or None (N) by inspecting library
documentation, source code, and package metadata. Python libraries were
additionally verified by installation and API introspection. A library scored
UNVERIFIED on a dimension --- rather than guessed --- is excluded from that
dimension's coverage denominator; \texttt{oqs} (Rust) is UNVERIFIED throughout
and excluded entirely, leaving seven libraries with primary-source evidence
for the coverage figures in Section~\ref{sec:gap_results}.

\begin{table}[h]
\centering
\caption{PQC Libraries Evaluated (State as of September 2026)}
\label{tab:libraries}
\begin{tabular}{llll}
\toprule
\# & Library & Language & Notes \\
\midrule
1 & liboqs-python   & Python     & OQS project Python binding \\
2 & cryptography (pyca) & Python & De-facto Python crypto library \\
3 & liboqs-js       & JavaScript & OQS project JS binding \\
4 & noble-post-quantum & JavaScript & Audited pure-JS PQC \\
5 & pqcrypto        & Rust       & \emph{Archived 2026-09-16; excluded} \\
6 & RustCrypto ml-*, x-wing & Rust & Pure Rust, trait-based \\
7 & oqs (Rust)      & Rust       & \emph{Not audited; UNVERIFIED} \\
8 & cloudflare/circl & Go        & Production-deployed \\
9 & Bouncy Castle   & Java       & Enterprise Java crypto, CMVP-certified \\
\bottomrule
\end{tabular}
\end{table}

\begin{table}[h]
\centering
\caption{Scoring Rubric --- Nine Externally-Anchored Dimensions}
\label{tab:rubric}
\small
\begin{tabular}{p{1.9cm} p{5.5cm}}
\toprule
Dimension & Full (F) requires --- source \\
\midrule
D1 KEM parameter set & ML-KEM-1024 selectable and functional --- CNSA~2.0 \\
D2 Signature parameter set & ML-DSA-87 selectable and functional --- CNSA~2.0 \\
D3 Stateful signing & LMS or XMSS available --- CNSA~2.0, SP~800-208 \\
D4 Conformance evidence & Testable or tested against NIST-issued ACVP vectors, with the CAVP/CMVP distinction stated --- CMVP \\
D5 Hybrid combiner & Built-in hybrid KEM with stated domain separation, one API call --- IETF hybrid draft \\
D6 Defaults & Default construction meets D1/D2 (n/a for libraries exposing primitives only) --- CNSA~2.0 \\
D7 Discovery \& inventory & Automated, codebase-agnostic cryptographic asset inventory --- TNO CADI \\
D8 Migration path & Tooling to upgrade existing classical keys/ciphertext --- TNO CADI, CNSA~2.0 timelines \\
D9 Protocol integration & At least two of: TLS config, X.509, JWT, envelope encryption --- CNSA~2.0 timelines \\
\bottomrule
\end{tabular}
\end{table}

\subsection{Results}
\label{sec:gap_results}

\begin{figure}[t]
  \centering
  \includegraphics[width=\columnwidth]{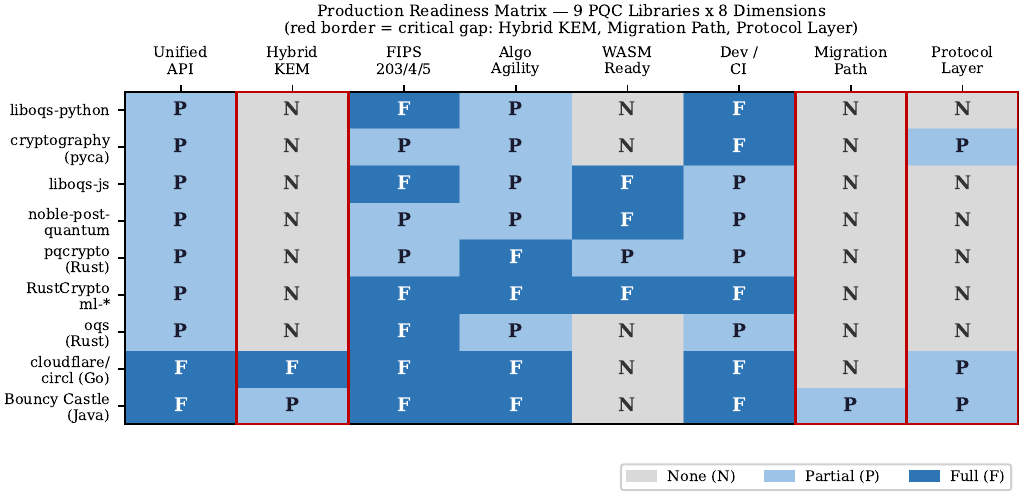}
  \caption{Production readiness matrix, nine externally-anchored dimensions
    (September 2026 audit). F~=~Full, P~=~Partial, N~=~None,
    U~=~UNVERIFIED (excluded from coverage figures). \qs{} scores below
    Bouncy Castle on D3 and D4, and Partial rather than Full on D6 against
    its own defaults.}
  \label{fig:gap_matrix}
\end{figure}

Fig.~\ref{fig:gap_matrix} shows the complete matrix. Table~\ref{tab:coverage}
summarises coverage per dimension across the seven libraries with
primary-source evidence; per-dimension $N$ varies where a library was
UNVERIFIED on that dimension specifically (noted in the table) rather than
where it was audited and scored None.

\begin{table}[h]
\centering
\caption{PQC Ecosystem Coverage by Dimension ($N$ verified libraries per row)}
\label{tab:coverage}
\begin{tabular}{lrrrl}
\toprule
Dimension & $N$ & Full & Partial & Coverage (F+P) \\
\midrule
D1 KEM parameter set    & 7 & 7 & 0 & 100\% \\
D2 Sig.\ parameter set  & 7 & 7 & 0 & 100\% \\
\textbf{D5 Hybrid combiner} & \textbf{6} & \textbf{2} & \textbf{1} & \textbf{50\%} \\
D9 Protocol integration & 7 & 2 & 1 & 43\% \\
D4 Conformance evidence & 5 & 1 & 1 & 40\% \\
\textbf{D3 Stateful signing} & \textbf{7} & \textbf{1} & \textbf{0} & \textbf{14\%} \\
D8 Migration path        & 7 & 0 & 1 & 14\% \\
\textbf{D7 Discovery \& inventory} & \textbf{7} & \textbf{0} & \textbf{0} & \textbf{0\%} \\
\bottomrule
\end{tabular}
\end{table}

\subsection{Critical Gaps and Per-Library Analysis}
\label{sec:gap_critical}

\textbf{Discovery and inventory (D7: 0\% across seven audited libraries).}
No library we could verify provides automated cryptographic asset discovery.
Every one exposes primitives without a way to inventory which of them a given
codebase uses, which the TNO CADI market survey~\cite{tnocadi2025} treats as a
requirement for migration tooling generally, not a nicety. \qs{} closes this
with a CycloneDX Cryptographic Bill of Materials (Section~\ref{sec:design}).

\textbf{Hybrid combiners (D5: 50\% of six verified, 33\% Full) --- revised
downward in rarity from an earlier finding.} noble-post-quantum ships preset
combiners (\texttt{ml\_kem768\_x25519} and others) with an HKDF-SHA256
combiner; RustCrypto's \texttt{x-wing} crate is a dedicated hybrid PQ KEM
(X25519\,+\,ML-KEM-768); cloudflare/circl provides \texttt{X25519MLKEM768()}
but states its combiners use ``simple concatenation'' and are ``not proven
secure in broader context'', scoped to TLS --- scored Partial here rather than
Full, since that is narrower than the domain-separation and key-binding
obligations D5 anchors to. Bouncy Castle's hybrid support could not be
confirmed from the sources consulted and is excluded from this dimension's
denominator rather than guessed. An earlier version of this evaluation, using
a self-selected rubric and a March-2026 snapshot, found only one library with
a combiner; that finding does not hold nine months later, and the ecosystem
moved rather than the earlier count being wrong at the time. IETF's hybrid
draft~\cite{ietf-hybrid-kem} still specifies precise domain separation and key
binding requirements that a manual concatenation does not by itself
satisfy~\cite{georgiev2012most}, so partial combiner availability is a
narrower gap than none, not a closed one.

\textbf{Stateful signing (D3: 14\%) and conformance evidence (D4: 40\%,
1 of 5 Full) --- Bouncy Castle outscores this library on both.} Bouncy
Castle is the only audited library supporting both LMS and XMSS
(SP~800-208), and the only one holding an active CMVP FIPS~140-3
certificate (\#4943) --- validation by an accredited laboratory, which no
self-testing produces. \qs{} adds LMS support and reports 225/225 runnable
ACVP cases (Section~\ref{sec:related_rubric} and
Section~\ref{sec:eval}); neither closes the gap to a CMVP certificate.
noble-post-quantum's README states it tests against ACVP and Wycheproof
vectors without claiming validation, which is the same distinction this paper
holds itself to.

\textbf{Migration path (D8: 14\%).} Adopting PQC is not only an API question:
existing deployments carry classical keypairs that must be upgraded and stored
ciphertext that may need re-encryption. Only Bouncy Castle (partially)
addresses this among audited libraries. The companion QSA
tool~\cite{shaw2025qsa} addresses the detection half of this problem for
Python codebases; \qs{} addresses the replacement half.

\textbf{Protocol layer (D9: 43\%).} pyca/cryptography now provides full
X.509 support alongside its newly-added ML-KEM/ML-DSA primitives (see
``Why Python'' below), and Bouncy Castle provides TLS and X.509/PKIX; circl
provides HPKE and PKI support scored Partial. This dimension is less severe
than the earlier snapshot found, driven by pyca/cryptography's addition of
X.509 support for the algorithms it added.

\subsection{Why Python}

Python runs on approximately 37\% of production web
services\footnote{Stack Overflow Developer Survey 2024.} and is the dominant
language in data science, ML inference, and API gateway segments where PQC
migration is most urgent. A gap in the Python PQC ecosystem therefore affects
a large fraction of production deployments.

That gap has narrowed since an earlier version of this work: pyca/cryptography
48.0.0 now ships \texttt{MLKEM768}/\texttt{MLKEM1024} and
\texttt{MLDSA44}/\texttt{65}/\texttt{87} natively, closing what was previously
the Python \emph{algorithm} gap. It provides no hybrid combiner, no migration
tooling, and no discovery capability, and liboqs-python provides primitives
only. Restated precisely: \textbf{in Python specifically, no library other than
\qs{} combines a hybrid KEM, migration tooling, and protocol helpers in one
package.} That is a narrower and more defensible claim than ``the ecosystem
lacks a production layer'', and it is the one the September~2026 audit
supports. It is the primary motivation for \qs{}.

\section{Library Design}
\label{sec:design}

\subsection{Design Principles}

The library is built around five explicit design principles, each traceable
to a specific failure mode observed in the gap analysis.

\textbf{P1: Hybrid by default.} Opting into classical-only mode requires an
explicit flag, inverting the usual burden: the easy path is the secure path,
rather than code that passes tests while silently lacking quantum
resistance.

\textbf{P2: Backend agnostic.} An abstract backend layer separates the API
from algorithm implementations; the same application code runs against
liboqs~\cite{liboqs} (default) or RustCrypto (PyO3) unmodified, protecting
against a backend becoming unavailable.

\textbf{P3: Protocol ready.} TLS configuration helpers, X.509 hybrid
certificate generation, CBOR envelope encryption, and JWT signing ship as
first-class interfaces; raw byte access remains available but secondary.

\textbf{P4: Migration first.} Key formats are versioned CBOR~\cite{rfc8949}
from v1.0. An upgrade module transforms classical keypairs to hybrid ones;
a scanner (\texttt{qs-audit}) locates classical-crypto imports, complementing
the companion QSA tool's~\cite{shaw2025qsa} deeper static analysis.

\textbf{P5: Safe defaults, with a stated limit.} The default is ML-KEM-768 /
Ed25519+ML-DSA-65 (level~3) rather than level~1. But CNSA~2.0~\cite{cnsa20}
mandates ML-KEM-1024/ML-DSA-87 (level~5, both supported, neither default), so
these defaults sit \emph{below} that suite --- scored as a shortfall in
Section~\ref{sec:related_rubric} rather than noted in passing. \texttt{qs-audit
cnsa2} reports or enforces the distinction.

\subsection{Module Architecture and API Usability}

The five design principles map directly to the module structure: application
code touches only \texttt{HybridKEM} and \texttt{HybridSign}; protocol
helpers (\texttt{tls}, \texttt{x509}, \texttt{envelope}) and the audit module
sit above a backend abstraction layer currently exposing liboqs-python and
RustCrypto. Adding a backend requires implementing one abstract base class
in \texttt{backends/base.py}, with no application-code changes.

The hybrid KEM task in \qs{} is three lines (Listing~\ref{lst:qs_api}). The
equivalent liboqs-python task needs separate X25519 and ML-KEM keygen, manual
key concatenation with no standard format, and an HKDF combination step whose
salt, domain-separation string, and share ordering are left to the
application to get right --- roughly 45 lines, and exactly the kind of
detail~\cite{georgiev2012most} finds developers get wrong. The
IETF~\cite{ietf-hybrid-kem} draft specifies answers to all three questions;
\qs{} implements the specification so the application does not have to.

\begin{lstlisting}[caption={Hybrid KEM in \qs{} --- three lines.},
  label=lst:qs_api]
from quantum_safe.kem.hybrid import HybridKEM
kem = HybridKEM()          # default: X25519 + ML-KEM-768
kp  = kem.generate_keypair()
ct, shared_secret = kem.encapsulate(kp.public)
shared_secret_b   = kem.decapsulate(kp, ct)
assert shared_secret == shared_secret_b
\end{lstlisting}

\subsection{Key Serialisation and Versioning}
\label{sec:design_serial}

Keys are serialised in two formats. PEM is provided for compatibility with
existing tooling. CBOR~\cite{rfc8949} is the primary format: it encodes
the algorithm identifier, parameter set, and key material in a compact
binary encoding with a version field that supports future migrations.
The CBOR schema is:

\begin{lstlisting}[language={},caption={CBOR hybrid key structure (CDDL notation).},
  label=lst:cbor]
hybrid-key = {
  "v"   : uint,          ; version (currently 1)
  "alg" : text,          ; e.g. "X25519+ML-KEM-768"
  "cls" : bstr,          ; classical component bytes
  "pqc" : bstr,          ; PQC component bytes
  ? "params" : map,      ; optional algorithm parameters
}
\end{lstlisting}

The version field serves the migration-first principle: when ML-KEM-768 is
eventually superseded, the library can detect legacy keys and apply an
automated upgrade path. This design prevents the silent failure mode where
an application continues to use a deprecated algorithm because no tooling
exists to detect or replace it.

\section{Benchmark Methodology}
\label{sec:methodology}

\subsection{Harness Design}

All benchmarks share a common harness. Table~\ref{tab:harness} documents every
design decision and its rationale.

\begin{table}[h]
\centering
\caption{Benchmark Harness Design Decisions}
\label{tab:harness}
\small
\begin{tabular}{p{2.0cm} p{1.8cm} p{3.8cm}}
\toprule
Parameter & Value & Rationale \\
\midrule
Iterations   & 3,000 per run  & 2,940 samples per run after the trim; raw samples archived \\
Warmup       & 100    & Eliminates JIT, cold-cache, and import effects \\
Outlier trim & 1\%    & Removes OS scheduling spikes; preserves distribution shape \\
Timer        & {\small\texttt{time.perf\_\newline{}counter}} & Nanosecond resolution; monotonic \\
GC           & Disabled & Prevents GC pauses skewing samples \\
CPU pinning  & cores 0--1 & Eliminates cross-core migration noise \\
Runs         & 1 warm-up (discarded) + 15 timed & Median across runs, never the fastest; intervals are bootstrap over runs \\
\bottomrule
\end{tabular}
\end{table}

The trim formula is $\texttt{samples}[\textit{clip}\,{:}\,N - \textit{clip}]$
where $\textit{clip} = \max(1, \lfloor N \times 0.01 \rfloor)$.
Python's garbage collector (\texttt{gc.disable()}) is disabled for the duration
of each measurement loop and re-enabled between operations.

\subsection{Statistical Measures}

We report five statistics for every operation. The \emph{median} (p50) is
the headline figure: for a right-skewed timing distribution with occasional
OS scheduling spikes, the median is a more stable estimator than the mean.
The 95th percentile (p95) characterises tail latency relevant to SLA design.
The Coefficient of Variation (\textit{CoV}\,=\,$\sigma/\mu \times 100$) is
reported per operation as a stability metric under repetition. It is computed
with the secret held fixed and does not by itself constrain secret dependence;
that question is addressed separately in a companion
paper~\cite{twoclass2027}.

For hypothesis testing, we compute Welch's two-sample $t$-test~\cite{welch1947}
rather than Student's $t$-test, because the classical and hybrid timing
distributions have different variances. The $t$-statistic is:
\begin{equation}
  t = \frac{\bar{x}_B - \bar{x}_A}{\sqrt{s_A^2/n_A + s_B^2/n_B}}
  \label{eq:welch_t}
\end{equation}
with Welch--Satterthwaite effective degrees of freedom:
\begin{equation}
  \nu = \frac{(s_A^2/n_A + s_B^2/n_B)^2}
             {(s_A^2/n_A)^2/(n_A-1) + (s_B^2/n_B)^2/(n_B-1)}
  \label{eq:satterthwaite}
\end{equation}
Effect size is reported as Cohen's $d$~\cite{cohen1988}:
\begin{equation}
  d = \frac{\bar{x}_B - \bar{x}_A}{s_p}, \quad
  s_p = \sqrt{\frac{(n_A-1)s_A^2 + (n_B-1)s_B^2}{n_A+n_B-2}}
  \label{eq:cohens_d}
\end{equation}

Bootstrap 95\% confidence intervals follow the percentile method of Efron~\cite{efron1979bootstrap} with $B = 10{,}000$ resamples of the
per-run medians (15 runs, the warm-up run discarded). They describe how far a reported median moves from run to run on this machine. We do not
bootstrap the samples inside a run, because they are serially correlated; for the same reason Welch's test and Cohen's $d$ are computed on
per-run values ($n_A = n_B = 15$), not on the roughly 2,940 samples per run, which would give a far larger and meaningless $t$. The raw samples
of every run are archived, and every statistic in this paper is computed from them by script.

\subsection{Test Environments}

Two environments, same physical hardware (Intel Core i9-14900HX, x86-64; 64-bit Windows~11 host on AC power, processor minimum state set to 100\%).
\textbf{ENV-2 (primary, authoritative for all paper claims):} Docker
(\texttt{python:3.12-slim}, Python~3.12.14), Linux 6.18.33.2-WSL2, liboqs~0.15.0 built from
source with \texttt{-DOQS\_DIST\_BUILD=ON} (AVX2/AVX-512 dispatch enabled),
CPU-pinned to cores~0--1. \textbf{ENV-1 (comparison):} Windows~11 native
Python~3.12.7, liboqs~0.15.0 via the generic MSYS2 DLL (no AVX2/AVX-512).
ENV-1 quantifies the build-flag effect and informs Windows deployment
decisions.

\subsection{Reproducibility}

Every number in this paper is computed from run files written by one script:

\begin{lstlisting}[language=bash]
bash tests/bench/run_env2.sh   # builds the image from the current commit, runs 1 warm-up + 15 timed
                               # runs, the leakage screen, and records the environment
python tests/bench/paper_numbers.py results/env2/<run-dir> \
    --env1 results/env1/<run-dir> --out paper_numbers.json
\end{lstlisting}

The benchmark scripts accept \texttt{--iterations} and \texttt{--raw-samples}; with the latter every timed sample is saved, next to the
machine, kernel, liboqs version and code commit. \texttt{paper\_numbers.py} produces the tables, the interval estimates and the figures'
inputs; nothing is typed by hand. The raw run files, the environment record and these scripts are archived with the paper.
Earlier versions of this work reported figures that this procedure supersedes (see the correction note after Table~\ref{tab:decomp}).
The statistical analysis module (\texttt{tests/bench/bench\_stats.py}) implements the five statistical measures described above.

\section{Evaluation}
\label{sec:eval}

\subsection{Hybrid KEM Overhead}
\label{sec:eval_kem}

\subsubsection{Decomposition}

\begin{figure}[t]
  \centering
  \includegraphics[width=\columnwidth]{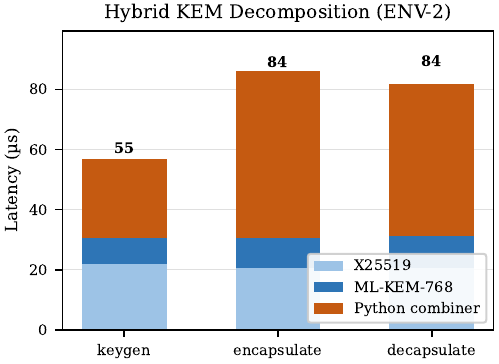}
  \caption{Stacked bar decomposition of hybrid KEM latency. The Python
    combiner (HKDF-SHA256, PEM/CBOR serialisation, key wrapping) accounts
    for 26.2--55.7\us{} per operation --- comparable to or exceeding the
    underlying cryptographic primitive cost in ENV-2.}
  \label{fig:decomp}
\end{figure}

Table~\ref{tab:decomp} and Fig.~\ref{fig:decomp} show the three-tier
decomposition: X25519 alone (tier~\circled{1}), ML-KEM-768 alone
(tier~\circled{2}), and the full hybrid combiner (tier~\circled{3}).
The combiner overhead is computed as $\text{\circled{3}} - \text{\circled{1}} - \text{\circled{2}}$.

\begin{table*}[!t]
\centering
\caption{Hybrid KEM Decomposition --- ENV-2 (Docker/Linux, 3{,}000 iterations per run). Median across 15 timed runs after one discarded warm-up run; range = fastest to slowest run; p95 and CoV are medians of the per-run values.}
\label{tab:decomp}
\small
\begin{tabular}{llrrrr}
\toprule
Tier & Operation & Median & Run range & p95 & CoV \\
\midrule
\multirow{2}{*}{\circled{1} X25519}
  & keygen         & 21.9\us & 17.8--25.9\us & 28.9\us & 16.6\% \\
  & DH exchange    & 20.6\us & 18.7--25.6\us & 25.9\us & 5.6\% \\
\midrule
\multirow{3}{*}{\circled{2} ML-KEM-768}
  & keygen         & 8.7\us & 8.6--11.9\us & 9.7\us & 6.3\% \\
  & encapsulate    & 9.9\us & 9.1--12.7\us & 12.6\us & 6.6\% \\
  & decapsulate    & 10.7\us & 10.3--14.5\us & 14.1\us & 9.8\% \\
\midrule
\multirow{3}{*}{\circled{3} HybridKEM}
  & keygen         & 55.1\us & 51.4--72.5\us & 76.5\us & 16.2\% \\
  & encapsulate    & 83.7\us & 75.4--103\us & 112\us & 14.7\% \\
  & decapsulate    & 84.2\us & 77.2--104\us & 118\us & 15.1\% \\
\midrule
\multirow{3}{*}{Combiner overhead}
  & keygen         & 26.2\us & 15.2--41.4\us & --- & --- \\
  & encapsulate    & 55.7\us & 38.7--75.1\us & --- & --- \\
  & decapsulate    & 50.3\us & 42.4--75.1\us & --- & --- \\
\midrule
\multicolumn{2}{l}{\textbf{Full handshake} (\circled{3} keygen + encap + decap, per run)}
  & \textbf{231\us} & 214--277\us & --- & --- \\
\bottomrule
\end{tabular}
\end{table*}

The full hybrid KEM handshake (keygen + encapsulate + decapsulate) takes a median of \textbf{231\us{}} in ENV-2 (95\%
interval 218--250\us{} over 15 runs; fastest run 214\us{}, slowest 277\us{}).
The combiner overhead is substantial: about 26.2\us{} on keygen and 50.3--55.7\us{} on encapsulate and decapsulate, together 57\% of the handshake
(95\% interval 54--62\%). By construction the overhead is everything other than the two
primitives, \circled{3}\,$-$\,\circled{1}\,$-$\,\circled{2}: Python serialisation, HKDF and object handling, not cryptography. We have not
measured an optimised path and make no estimate of the saving a cache of serialised keys would give.

\textbf{Correction note.} Earlier versions of this work reported 243\us{} for this handshake and tabulated ML-KEM-768 decapsulation at
12.55\us{}. Three defects produced those numbers. The harness shared one liboqs object across the keygen, encapsulation and decapsulation
measurements, so decapsulation ran against a key that did not match its ciphertext and exercised the FIPS~203 implicit-rejection path. The
tables reported the fastest of three runs, which is a minimum order statistic. And the \texttt{--iterations} flag was parsed but ignored,
so tables captioned ``3,000 iterations'' were measured with 1,000. All three are fixed and the figures above supersede the earlier ones. The
change is not a uniform shift (hybrid keygen is now 55.1\us{} where 100\us{} was reported, hybrid encapsulate and decapsulate
are 83.7--84.2\us{} where 72\us{} was reported), and absolute figures differ between measurement sessions on
the same host, so the earlier numbers should not be mixed with these.

The X25519-only handshake in the same library (keygen plus one DH exchange) takes a median of 43.5\us{}, so the hybrid handshake
is 5.3$\times$ slower (95\% interval 4.7--5.8$\times$). Welch's $t$-test
(Eq.~\eqref{eq:welch_t}) on the 15 per-run handshake totals gives $t \approx 38.4$, $\nu \approx 16.5$,
$p \approx 1.5\times10^{-17}$ and Cohen's $d \approx 14.0$ --- unambiguous, as expected for a five-fold increase, though the relevant
question is operational acceptability, not detectability. Against a LAN/intra-datacenter TLS~1.3 budget of
8--40\ms~\cite{sosnowski2025layered,stebila2021pqtls}, the 184\us{} hybrid overhead (hybrid minus X25519-only) is
0.46--2.3\% of that budget, and 0.18\% at WAN latency (100\ms{} RTT)
--- negligible relative to any realistic network latency.

\subsection{Signature Overhead}
\label{sec:eval_sig}

Table~\ref{tab:sigs} reports signature benchmarks. The full hybrid signature
operation (HybridSign keygen + sign + verify) takes a median of
$\mathbf{364\us{}}$ (95\% interval 358--374\us{}).

\begin{table*}[!t]
\centering
\caption{Signature Benchmarks --- ENV-2 (Docker/Linux, 3{,}000 iterations per run; median across 15 runs, same conventions as Table~\ref{tab:decomp})}
\label{tab:sigs}
\small
\begin{tabular}{llrrrr}
\toprule
Suite & Operation & Median & Run range & p95 & CoV \\
\midrule
\multirow{2}{*}{Ed25519 (classical)}
  & sign (32\,B)    & 24.7\us & 24.2--27.2\us & 27.2\us & 5.3\% \\
  & verify (32\,B)  & 79.4\us & 76.6--82.1\us & 87.7\us & 5.2\% \\
\midrule
\multirow{3}{*}{ML-DSA-65}
  & keygen          & 28.1\us & 25.6--36.1\us & 35.2\us & 10.8\% \\
  & sign (32\,B)    & 61.6\us & 55.2--81.3\us & 158\us & \textbf{54.0\%} \\
  & verify (32\,B)  & 32.1\us & 24.9--35.4\us & 38.1\us & 12.9\% \\
\midrule
\multirow{3}{*}{HybridSign}
  & keygen          & 96.0\us & 85.8--100\us & 116\us & 14.2\% \\
  & sign (32\,B)    & 150\us & 140--172\us & 270\us & 32.7\% \\
  & verify (32\,B)  & 115\us & 111--157\us & 156\us & 12.2\% \\
\midrule
\multirow{2}{*}{X.509 Hybrid}
  & HybridCert build & 238\us & 219--276\us & 371\us & 22.5\% \\
  & verify\_cosig   & 209\us & 195--262\us & 282\us & 14.9\% \\
\bottomrule
\end{tabular}
\end{table*}

The large CoV for ML-DSA-65 sign (54\%) and its p95/median ratio of 2.6
reflect the rejection sampling loop in FIPS~204~\cite{fips204} \S5.2.
This is discussed in detail in Section~\ref{sec:eval_cov}.

\subsection{Concurrent Load Throughput}
\label{sec:eval_concurrent}

\begin{figure}[t]
  \centering
  \includegraphics[width=\columnwidth]{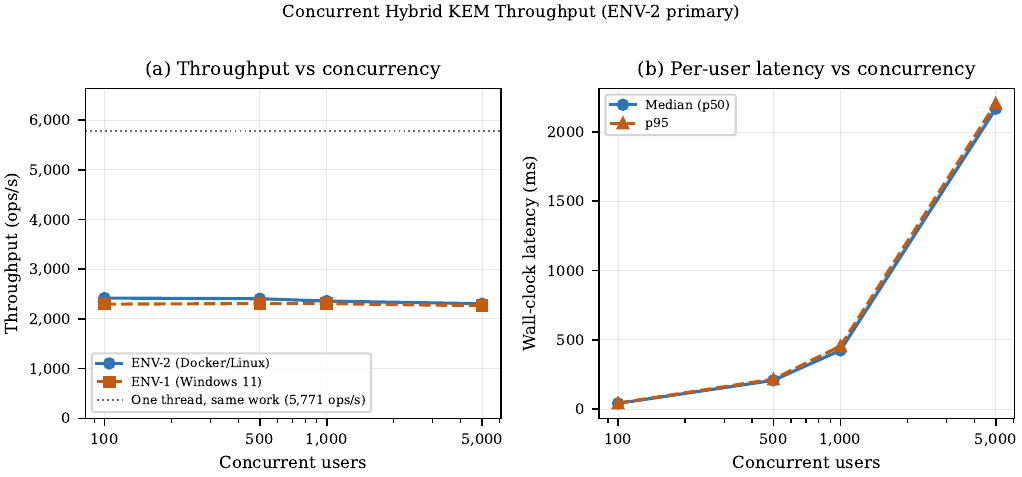}
  \caption{(a) Throughput vs concurrency; (b) per-user latency vs concurrency.
    ENV-2 and ENV-1 show nearly identical throughput despite a 1.9$\times$
    per-operation latency difference; both stay well below the rate of a single thread (dotted line).}
  \label{fig:throughput}
\end{figure}

Table~\ref{tab:concurrent} and Fig.~\ref{fig:throughput} show concurrent throughput from 100 to 5,000 simultaneous threads. Each thread
(a ``user'') performs one encapsulation and one decapsulation, and the reported time is the wall-clock time for all users.

\begin{table*}[!t]
\centering
\caption{Concurrent Hybrid KEM Throughput --- ENV-2 (Docker/Linux) and ENV-1 (Windows~11). One ``user'' is one thread doing one encapsulation and one decapsulation; time is wall-clock for all users; median across 15 runs.}
\label{tab:concurrent}
\small
\begin{tabular}{lrrrr}
\toprule
Users & Med.\ latency & p95 & Throughput & CoV \\
\midrule
\multicolumn{5}{l}{\emph{ENV-2 (Docker/Linux)}} \\
100   & 41.3\ms & 43.5\ms & 2,420 ops/s & 2.7\% \\
500   & 207.8\ms & 216.3\ms & 2,406 ops/s & 1.5\% \\
1,000 & 423.9\ms & 455.2\ms & 2,359 ops/s & 3.1\% \\
5,000 & 2,168.3\ms & 2,207.0\ms & 2,306 ops/s & 0.9\% \\
\midrule
\multicolumn{5}{l}{\emph{ENV-1 (Windows 11)}} \\
100   & 43.5\ms & 46.0\ms & 2,298 ops/s & 2.7\% \\
500   & 216.5\ms & 223.5\ms & 2,309 ops/s & 1.2\% \\
1,000 & 433.4\ms & 447.9\ms & 2,308 ops/s & 1.1\% \\
5,000 & 2,207.0\ms & 2,225.1\ms & 2,266 ops/s & 0.6\% \\
\bottomrule
\end{tabular}
\end{table*}

Throughput is flat: 2,420~ops/s at 100 users and 2,306~ops/s at 5,000 (4.7\% lower) in
ENV-2, and 2,298 and 2,266~ops/s in ENV-1. Flat throughput shows that the system degrades
gracefully under load; it does not show parallelism. A single thread running the same two operations back to back reaches about
5,771~users/s (95\% interval 5,489--6,041), so the threaded runs achieve
40\% of the serial rate: the threads add overhead rather than speed. The same holds in ENV-1,
where the per-operation cost is 1.9$\times$ higher but throughput is the same, which is what a limit set by Python's thread handling,
not by the cryptography, predicts.

\textbf{Correction.} An earlier version read the flat curve as evidence that liboqs releases the GIL, by comparison with a ``serialised''
estimate of 0.8~ops/s. That estimate inverted the quantity: serial execution gives about $1/243\us{} \approx 4{,}100$~ops/s, not 0.8, and the
observed throughput is below it. The inference does not hold, and we withdraw it.

\textbf{Testing the GIL question directly.} Table~\ref{tab:gil} splits a fixed amount of work across 1, 2 and 4 threads (a container with
4 CPUs) and reports the speedup over one thread. A pure-Python loop, which holds the GIL, is the negative control and shows 1.09$\times$
on four threads, as it should. Raw liboqs ML-DSA-65 signing (about 75\us{} per call) scales 1.6$\times$ on two threads and
2.6$\times$ on four, so the binding does release the GIL while the C code runs. The much shorter ML-KEM-768 decapsulation (about
13\us{}) scales 1.6$\times$ on two threads but is slower than one thread on four (0.73$\times$): handing the lock
back and forth costs about as much as the call. The Python layer around the primitives holds the GIL, and HybridKEM encapsulate-plus-decapsulate
gets slower with threads (0.62$\times$ on two, 0.54$\times$ on four). A server that wants several cores for hybrid KEM
should use processes, not threads.

\begin{table}[h]
\centering
\caption{GIL scaling test, ENV-2 container with 4 CPUs: speedup over one thread for the same total work (median of 9 repeats; a second full run gave the same pattern). 1.0 means no benefit from threads.}
\label{tab:gil}
\small
\begin{tabular}{lrrr}
\toprule
Workload & 1 thread & 2 threads & 4 threads \\
\midrule
Pure-Python loop (control) & 209,012 ops/s & 1.07$\times$ & 1.09$\times$ \\
liboqs ML-KEM-768 decap & 78,787 ops/s & 1.62$\times$ & 0.73$\times$ \\
liboqs ML-DSA-65 sign & 13,256 ops/s & 1.62$\times$ & 2.59$\times$ \\
HybridKEM encap+decap & 5,495 ops/s & 0.62$\times$ & 0.54$\times$ \\
\bottomrule
\end{tabular}
\end{table}

\subsection{Timing Stability and Cross-Environment Summary}
\label{sec:eval_cov}

CoV ($\sigma/\mu \times 100\%$ of the 1\%-trimmed sample) is reported here as
a stability metric, not a secret-dependence test: only the keygen
measurements draw fresh key material per iteration, so CoV describes how
stable an operation is under repetition with the secret fixed, not whether
timing depends on the secret. That question, and the two-class methodology
it actually requires, is the subject of the companion
paper~\cite{twoclass2027}; we summarise only the headline figures here.
Table~\ref{tab:cov} reports CoV against an AES-256-GCM noise floor of 3.0\%
in ENV-2 (median of per-run values). ML-KEM-768 operations sit 3--7 percentage points above that floor (6.3--9.8\%), and the HybridKEM
operations are higher still (15--16\%), which we attribute to the Python layer and the scheduler of this environment
but have not isolated. CoV is a stability description, not a pass/fail test. ML-DSA-65 signing (54\%) is high by FIPS~204's
rejection-sampling design, not by defect --- the companion paper measures this directly rather than arguing it from the specification.

\begin{table}[h]
\centering
\caption{Timing Stability --- ENV-2 CoV Reference Table (median of per-run CoV)}
\label{tab:cov}
\small
\begin{tabular}{lrl}
\toprule
Operation & CoV & Note \\
\midrule
AES-256-GCM enc 1\,KB (ref.)       & 3.0\% & Noise floor \\
HKDF-SHA256                        & 3.6\% & Near the floor \\
Ed25519 verify                     & 5.4\% & Near the floor \\
ML-KEM-768 keygen / encap          & 6.3--6.6\% & Above the floor \\
ML-KEM-768 decap                   & 9.8\% & Above the floor \\
HybridKEM keygen / encap / decap   & 14.7--16.2\% & Python layer, scheduler \\
X.509 HybridCert build             & 22.5\% & Contains ML-DSA sign \\
HybridSign sign                    & 32.7\% & Expected (FIPS 204) \\
ML-DSA-65 sign                     & \textbf{54.0\%} & Expected (FIPS 204) \\
\bottomrule
\end{tabular}
\end{table}

Percentile profiling (p50/p95/p99) shows the ML-KEM-768 p95 at 13--32\% above the median and the HybridKEM p95 at 34--40\%; the wider
HybridKEM tails are consistent with the Python layer rather than the underlying primitive, which we did not isolate.

Table~\ref{tab:cross} compares ENV-1 and ENV-2 on identical hardware.

\begin{table}[h]
\centering
\caption{Cross-Environment Comparison --- Same Hardware, Different OS and liboqs Builds (medians)}
\label{tab:cross}
\small
\begin{tabular}{lrrr}
\toprule
Operation & ENV-1 & ENV-2 & Ratio \\
\midrule
ML-KEM-768 keygen      & 55.7\us & 8.7\us & 6.4$\times$ \\
ML-KEM-768 decap       & 24.7\us & 10.7\us & 2.3$\times$ \\
HybridKEM keygen       & 135\us & 55.1\us & 2.5$\times$ \\
Full KEM handshake     & 444\us & 231\us & 1.9$\times$ \\
ML-DSA-65 sign         & 261\us & 61.6\us & 4.2$\times$ \\
Throughput @ 5,000 u  & 2,266 ops/s & 2,306 ops/s & 1.02$\times$ \\
\bottomrule
\end{tabular}
\end{table}

ENV-2 is faster on every operation, by 1.5$\times$ to 6.4$\times$ (ML-KEM-768 keygen 6.4$\times$). The size of the raw ML-KEM
difference is consistent with the build difference (CPUID-gated AVX2/AVX-512 dispatch with \texttt{-DOQS\_DIST\_BUILD=ON} in ENV-2, a generic
MSYS2 DLL in ENV-1), but this comparison does not isolate it: the operating system, the timer and the container also differ, and we did not
rebuild either environment with the other's flags. We report the difference and the build as co-varying. Throughput at 5,000 users is the
same in both environments (2,266 and 2,306~ops/s), consistent with the limit set by Python's
thread handling described above.

\section{Discussion}
\label{sec:discussion}

\subsection{Limitations}

\textbf{Single physical host.} All benchmarks were collected on one machine.
Absolute latency figures are hardware-specific, and they also moved between measurement sessions on this host (hybrid keygen was 100\us{} in an earlier session and is 55.1\us{} here), so we report ranges and archive raw samples and do not claim that the absolute values transfer. The relative relationships ---
combiner overhead fraction, ENV-1/ENV-2 speedup ratio, concurrency scaling ---
are expected to generalise to comparable AMD64 hardware.

\textbf{CoV does not measure secret dependence.} Our CoV figures are computed
with a single key held fixed across the measurement, so they describe stability
under repetition rather than any relationship between the secret and execution
time. Answering the latter requires two measurement classes that differ in the
secret --- the fixed-versus-random design of dudect~\cite{reparaz2017dudect} ---
which we treat as a separate experiment rather than an interpretation of these
numbers. We caution specifically that a fixed-versus-random comparison
implemented at this level of abstraction is not self-interpreting: unless the
two classes are matched in allocation and construction, the class order is
randomised, and a random-versus-random control is run alongside the usual
fixed-versus-fixed one, the test reports a stable, significant $t$-statistic that
tracks byte-identical repetition rather than key material. A formal
constant-time proof requires tools like ct-verif~\cite{almeida2016cverif} or
dudect applied to the compiled implementation.

\textbf{Intervals describe run-to-run variation on one host.} The bootstrap intervals are over 15 runs in one session on one machine. They do not
include variation between machines, between sessions or between kernel versions.

\textbf{WASM target not yet released.} Principle P2 (backend agnostic)
is partially realised: the liboqs and RustCrypto backends are functional,
but the WASM target for browser deployment has not been published. This
is planned for version~0.2.

\textbf{FIPS~205 exposure, and a correction.} SLH-DSA is now reachable through
the liboqs backend across the SHAKE and SHA2 pure parameter sets. This corrects a
defect worth recording: the canonical \texttt{SLH-DSA-*} names previously
resolved to liboqs' \texttt{SPHINCS+-...-simple} mechanisms, which are the
pre-standard round-3 submission rather than FIPS~205. The two share parameter
sizes, so the substitution was invisible from key and signature lengths, but they
are not interchangeable --- FIPS~205 prepends a domain-separation prefix, and we
confirm by cross-verification that a signature produced under one does not verify
under the other. Any earlier claim of FIPS~205 support in this library should be
read as pre-standard SPHINCS+.

\textbf{Stateful hash-based signing: LMS added, XMSS still absent.} CNSA~2.0
requires LMS or XMSS (SP~800-208) for software and firmware signing; neither
is available in the liboqs backend. Because a stateful scheme fails
catastrophically on one-time-key reuse --- we confirmed directly that
deserialising a stale key snapshot rewinds its index and that two different
messages signed at the rewound index both verify, which is enough to forge
further signatures --- we regard wrapping an existing, carefully-maintained
implementation as preferable to writing one. \qs{} adds LMS (RFC~8554) via an
optional dependency (\texttt{pyhsslms}, maintained by an RFC~8554 co-author),
with a wrapper that write-ahead reserves each index through a caller-supplied
durable store before releasing a signature and refuses a key presented at an
already-issued index. This satisfies D3 as Partial, not Full: XMSS remains
absent, and SP~800-208 additionally requires key generation inside a validated
module, which the library cannot provide for the caller. Bouncy Castle, the
only audited library scoring Full on D3, supports both schemes and holds the
CMVP certificate this dimension's Full tier is ultimately anchored to.

\textbf{Gap analysis scores.} The production-readiness scores are derived from
public documentation, source code, and package metadata. Library maintainers
may disagree with specific scores, particularly in the Partial category where
the boundary with None and Full involves judgment. All scoring decisions are
documented and publicly reproducible.

\subsection{Practical Deployment Implications}

A server handling 2,306 new hybrid sessions per second spends about 0.4 CPU core on the handshake overhead above
classical-only operation (184\us{} per session) --- acceptable for most applications. The 444\us{}
ENV-1 figure is consistent with a build effect rather than a Windows penalty; running in Docker or building liboqs from source with
\texttt{-DOQS\_DIST\_BUILD=ON} should bring a Windows host close to the ENV-2 figure, though we did not test a source build on Windows. \texttt{qs-audit} integrates into CI/CD to flag
new classical-crypto imports as they are introduced, and combined with the
companion QSA tool~\cite{shaw2025qsa} gives continuous migration-status
visibility.

This work was conducted without institutional affiliation; the benchmark
methodology, library code, tests, and scripts are released in their
entirety, and we rely on that reproducibility for credibility rather than
institutional authority. All gap-analysis scores are derived from public
documentation, source code, and package metadata, current as of the audit
date stated in Section~\ref{sec:gap_method}; readers can reproduce the
evaluation from the same sources.

\section{Related Work}
\label{sec:related}

\textbf{PQC performance evaluation.} Most prior work targets algorithm-level
benchmarks on embedded hardware (pqm4~\cite{kannwischer2019pqm4}, ARM
Cortex-M4; PQClean~\cite{pqclean}, reference implementations) or network-level
TLS effects on C-based implementations~\cite{paquin2020pqtls,stebila2021pqtls}.
None addresses hybrid construction, Python bindings, or concurrent server
throughput. Sosnowski et al.~\cite{sosnowski2025layered} find the TLS
handshake layer algorithm-neutral; our 231\us{} figure is consistent with
that.

\textbf{Library surveys.} Nejatollahi et al.~\cite{nejatollahi2019survey}
survey lattice-based implementations on hardware accelerators.
\cite{pqclibsurvey2025} surveys nine libraries for Kyber/Dilithium/FALCON/
SPHINCS+ support --- algorithm support, not production layer; only Bouncy
Castle overlaps our set. We do not claim no multi-library evaluation exists;
we claim none evaluates the production layer.

\textbf{Positioning against a self-selected rubric.}
\label{sec:related_rubric}
A matrix whose
dimensions are chosen by the authors of one of the entries invites the
objection that the result is determined before any library is examined.
Section~\ref{sec:gap_method} anchors every dimension to an external
requirement; three self-selected dimensions from an earlier version could
not be so anchored and are dropped, and the exercise scores \qs{} below its
own earlier claims on two dimensions (Section~\ref{sec:gap_critical}) --- a
result the audit produced, not a limitation we note in passing.

\textbf{API usability.} Georgiev et al.~\cite{georgiev2012most} and Lazar et
al.~\cite{lazar2014why} (269 CVEs) establish that API design predicts
cryptographic misuse. Our LOC comparison (Section~\ref{sec:design}) extends
this to hybrid PQC, arguing hybrid-by-default structurally prevents the most
common failure mode.

\textbf{Timing side-channel detection.} dudect~\cite{reparaz2017dudect} and
ct-verif~\cite{almeida2016cverif} operate at the binary or hardware level.
The CoV stability metric reported in this paper (Section~\ref{sec:eval_cov})
is not a substitute for either; the companion paper~\cite{twoclass2027}
develops the methodology that actually occupies their design space at the
Python-library level.

\textbf{Standards and migration tooling.} Ounsworth et
al.~\cite{composite-sigs} standardise composite signatures for Internet PKI,
which the \qs{} X.509 hybrid certificate implementation follows --- the first
Python implementation of the draft, to our knowledge. The companion
Quantum-Safe Auditor~\cite{shaw2025qsa} statically detects classical
cryptography in Python codebases (P~=~71.98\%, R~=~100\%) and assesses
migration urgency; together with \qs{}, detection and replacement form one
toolchain.

\section{Conclusion}
\label{sec:conclusion}

The PQC algorithm standardisation problem was solved in August 2024. The
production deployment problem is not, and a matrix that scores it must not
choose its own scoring dimensions. Anchoring nine dimensions to CNSA~2.0,
CMVP, the TNO CADI survey, and the IETF hybrid draft, rather than to \qs{}'s
own design, overturns one of this paper's own earlier findings --- hybrid
combiners are now present in three of six libraries we could verify, not one
--- and scores \qs{} below Bouncy Castle on stateful signing and conformance
evidence, and below its own CNSA~2.0 parameter sets on its own defaults. We
report all three rather than the version of the audit that would have flattered
the library it accompanies.

The dimension every audited library still fails is discovery: zero of seven
provide automated cryptographic asset inventory. \qs{} closes it with a
CycloneDX Cryptographic Bill of Materials, adds LMS (RFC~8554) with
write-ahead index reservation against one-time-key reuse, and reports 225 of
225 runnable ACVP cases against NIST-published vectors --- conformance
evidence, stated throughout as distinct from the CMVP validation Bouncy
Castle actually holds.

A full X25519\,+\,ML-KEM-768 handshake takes a median of 231\us{} under Docker/Linux ---
0.46--2.3\% of a typical TLS~1.3 round-trip budget. Throughput is flat from 100 to 5,000
concurrent threads but below the single-thread rate: threads do not speed up the hybrid path, although liboqs does release the GIL during
longer calls. The timing side-channel question is
treated in a companion paper~\cite{twoclass2027}, having found that the
dispersion metric this paper's earlier version proposed for it cannot answer
that question, and that the test which can requires a control this work
identifies.

The library, 932 tests, the conformance runner, the leakage harness, and the
rubric's verification procedure are released as open-source software under the
Apache~2.0 licence. Every benchmark number in this paper is computed by script from archived run files, and every rubric score is independently re-auditable against
the published procedure --- including the two that count against \qs{}
itself. The counts and the code described here are those of release 0.3.2;
a signature-verification flaw present in releases up to 0.3.0 was fixed in
0.3.1 (GitHub advisory GHSA-wqv6-gm9x-69x8). It concerns how a verifier reads
the hedging prefix, not the conformance or timing results reported here. Combined with the Quantum-Safe Auditor~\cite{shaw2025qsa} for
detecting classical cryptography in codebases, the tools presented here and in
the companion paper provide a Python-native PQC migration toolchain whose
remaining gaps, on the evidence in this paper, are stated rather than
assumed closed.


\bibliographystyle{IEEEtran}
\bibliography{references}

\end{document}